\documentclass{ifacconf}

\usepackage{graphicx}      % include this line if your document contains figures
\usepackage{natbib}        % required for bibliography
\usepackage{color}
\usepackage{amsmath,mathrsfs}
\usepackage{amsfonts}

\usepackage{epstopdf}    % include this line if your document contains figures
\usepackage{algorithm}
\usepackage{balance} 
\usepackage{comment}
\usepackage{algorithmicx}
\usepackage{algpseudocode}
\usepackage{xfrac}
\usepackage{bm}
\usepackage{pgfplots}
\usepgflibrary{shapes}
\usepackage{tikz}
\usepackage{verbatim}
\usetikzlibrary{arrows,calc,patterns}
\tikzstyle{Rect}=[draw=blue,line width=0.001pt,preaction={clip, postaction={pattern=north east lines, pattern color=blue,line width=0.1pt}}]
\tikzset{
	>=stealth',
	help lines/.style={dashed, thick},
	axis/.style={<->},
	important line/.style={thick},
	connection/.style={thick, dotted},
}

\usepackage{eso-pic}
\AddToShipoutPictureBG*{%
	\AtPageUpperLeft{%
		\setlength\unitlength{1in}%
		\hspace*{\dimexpr0.5\paperwidth\relax}
		\makebox(0,-1.75)[c]{
			\begin{tabular}{c c}
				Rodrigo A. Gonz\'alez et al.,
				An EM Algorithm for Lebesgue-sampled State-space Continuous-time System Identification. \\
				To appear in
				{\em 22nd IFAC World Congress},
				Yokohama, Japan, 2023
				uploaded to ArXiv \today \\
\end{tabular}}}}

\begin{document}
\begin{frontmatter}
\title{An EM Algorithm for Lebesgue-sampled State-space Continuous-time System Identification\thanksref{footnoteinfo}}

\thanks[footnoteinfo]{This work was supported by the Swedish Research Council under contract number 2016-06079 (NewLEADS), by the Digital Futures project EXTREMUM, by the Chilean National Agency for Research and Development (ANID) Scholarship Program/Doctorado Nacional/2020-21202410 and by the grants ANID-Fondecyt 1211630, ANID-Basal Project FB0008 (AC3E).}

\author[First]{Rodrigo A. Gonz\'alez} 
\author[Second,Third]{Angel L. Cede\~no} 
\author[Third]{Mar\'ia Coronel}
\author[Second,Third]{Juan C. Ag\"uero}
\author[Fourth]{Cristian R. Rojas}

\address[First]{Department of Mechanical Engineering, Eindhoven University of Technology, Eindhoven, The Netherlands}
\address[Second]{Electronic Engineering Department, Universidad T\'ecnica Federico Santa Mar\'ia, Valpara\'iso, Chile}
\address[Third]{Advanced Center for Electrical and Electronic Engineering, AC3E, Valparaíso, Chile}
\address[Fourth]{Division of Decision and Control Systems, KTH Royal Institute of Technology, Stockholm, Sweden}

\begin{abstract}                % Abstract of not more than 250 words.
This paper concerns the identification of continuous-time systems in state-space form that are subject to Lebesgue sampling. Contrary to equidistant (Riemann) sampling, Lebesgue sampling consists of taking measurements of a continuous-time signal whenever it crosses fixed and regularly partitioned thresholds. The knowledge of the intersample behavior of the output data is exploited in this work to derive an expectation-maximization (EM) algorithm for parameter estimation of the state-space and noise covariance matrices. For this purpose, we use the incremental discrete-time equivalent of the system, which leads to EM iterations of the continuous-time state-space matrices that can be computed by standard filtering and smoothing procedures. The effectiveness of the identification method is tested via Monte Carlo simulations.
\end{abstract}


\end{frontmatter}
%===============================================================================

\section{Introduction}
In digital control system design, signals are usually sampled equidistantly in time.
This approach has led to a well established system theory and a vast number of successful applications due to its analytical tractability \citep{astroem1984computer}. An alternative sampling scheme, called event-based sampling, consists in retrieving measurements based on the occurrence of an event rather than the passing of time instants. One of the most popular event-based sampling schemes is \textit{Lebesgue sampling}, and it consists of sampling a signal whenever it crosses fixed and regularly-partitioned thresholds.

First steps in modern Lebesgue sampling theory can be found in \cite{aastrom1999comparison}, where a comparison between first-order systems using periodic and event-based sampling was made. The take-away message is that the Lebesgue sampling scheme requires fewer measurements on average than the equidistant (Riemann) sampling scheme, which is natural since the sampling is done at arguably the most relevant instants. Another advantage of Lebesgue sampling is that, in principle, only 1 bit is needed to indicate that the signal has crossed a threshold, which can be of interest when dealing with communication network systems. 

One of the issues of standard Lebesgue sampling is that incorporating continuous-time noise in the model description leads to theoretical difficulties when defining the sampling time instants. This issue is overcome in this paper by considering a send-on-delta sampling strategy \citep{miskowicz2006send} that includes a quantizer with hysteresis. Such type of sampling has been explored by, e.g., \cite{kofman2006level}, and in addition to its theoretical advantages, it is convenient for implementing 1-bit coding communication and minimizing spurious sampling. 

In this work, we study how to identify continuous-time systems in state-space using continuous-time input data and Lebesgue-sampled output data. Although this framework resembles the discrete-time identification problem with quantized data, key differences can be observed in the continuous-time treatment of the noise, and the hysteresis effect of the quantization step. Some contributions in such framework can be found in, e.g., \cite{Gustafsson2009} and \cite{Bottegal2017}. In \cite{kawaguchi2016system}, transfer function identification under Lebesgue sampling was studied. This work considered an \textit{approximate} Lebesgue sampling scheme, which has the shortcoming that the output data are not threshold values, but real numbers. Thus, it is not suitable for scenarios with very limited communication bandwidth or computational resources. Other contributions \citep{sanchez2019identification} has focused on the design of experiments for transfer function identification in closed-loop with an event-based sampling scheme. These methods tune a controller so that the system enters into a limit cycle, facilitating the estimation. In contrast to these procedures, our approach admits any order for the continuous-time system, does not require controller tuning, is suited for open-loop identification, and can be performed on any input and output data set.

This work overcomes the shortcomings detailed above by providing an EM-based algorithm for identifying systems subject to Lebesgue sampling. In summary,
\begin{itemize}
	\item
	We present closed-form expressions for the E and M-steps of the EM algorithm tailored for the identification of Lebesgue-sampled linear continuous-time systems described in state-space form. This algorithm delivers maximum likelihood estimates at convergence under mild conditions. Two forms are discussed, which use discrete-time equivalents of the system in shift operator and delta operator forms. The discrete-time equivalents rely on a \textit{user-defined} sampling period $\Delta$ that may improve the estimation accuracy (for small $\Delta$) at the cost of a higher computational cost;
    \item We provide two alternatives for computing the filtering and smoothing procedures required for the EM iterations; and		
	\item
	We show the effectiveness of the proposed method via extensive Monte Carlo simulations.
\end{itemize}
The rest of the paper is organized as follows. In Section \ref{sec:problemformulation} the problem statement is described, and the EM algorithm for Lebesgue-sampled system identification is derived in Section \ref{sec:lebesguesampledsysid}. Section \ref{sec:simulations} illustrates the method with a numerical example, and conclusions are drawn in Section \ref{sec:conclusions}. Proofs of the main results can be found in the Appendix.

\textit{Notation}: All matrices and vectors are written in bold, and column vectors are utilized, unless transposed. We employ the notation $\{f(t_l)\}_{l=1}^M$ to denote the set of evaluations $\{f(t_1),f(t_2),\dots,f(t_M)\}$, and $\{x(t)\}_{t\in[t_1,t_M]}$ to denote the continuous-time signal defined on the closed interval $[t_1,t_M]$.  A discrete-time signal is also written as $\{f_k\}_{k=1}^M$, depending on the context. The notation $\mathbf{x}_{1:N}$ describes $\{\mathbf{x}_k\}_{k=1}^N$. The Kronecker delta function is denoted as $\delta^K_k$, and the Dirac delta distribution is written as $\delta(t)$.

\section{Problem formulation}
\label{sec:problemformulation}
We consider the following linear time-invariant (LTI), single-input, single-output, continuous-time model:
\begin{subequations}
	\label{system2}
	\begin{align}
	\label{system2a}
	\dot{\mathbf{x}}(t) &= \mathbf{A}(\bm{\theta})\mathbf{x}(t) + \mathbf{B}(\bm{\theta})u(t) + \dot{\mathbf{w}}(t), \\
	\label{system2b}
	z(t) &= \mathbf{C}(\bm{\theta})\mathbf{x}(t) + D(\bm{\theta})u(t),
	\end{align}
\end{subequations}
where $\mathbf{A}(\bm{\theta}),\mathbf{B}(\bm{\theta}),\mathbf{C}(\bm{\theta})$, and $D(\bm{\theta})$ are $\bm{\theta}$-dependent matrices of suitable dimensions, and $\dot{\mathbf{w}}(t)$ is the formal derivative of a Wiener process of finite incremental covariance $\mathbf{Q}(\bm{\theta})$. The initial condition $\mathbf{x}(0)$ is assumed to be Gaussian-distributed with mean $\bm{\mu}_1$ and covariance $\mathbf{P}_1$, and the continuous-time white noise $\dot{\mathbf{w}}(t)$ is also assumed Gaussian with zero mean. %Note that the system is modeled by continuous-time dynamics in the state equation driven by continuous-time white noise and a deterministic~input.

We now introduce the sampling scheme of interest in this paper, which is also carefully explained by \cite{kofman2006level}. Given $\tau>0$ and the continuous-time output $z(t)\colon \mathbb{R}\to \mathbb{R}$, we define the sampled sequence $\{y(t_l)\}_{l=0}^\infty$ with quantization interval $h$ by the piecewise constant function $y(t)\colon \mathbb{R}\to \mathbb{R}$ that satisfies
\begin{equation}
\label{Qh}
y(\hspace{-0.01cm}t\hspace{-0.01cm}) \hspace{-0.09cm}= \hspace{-0.08cm} \mathcal{Q}_\tau\hspace{-0.02cm}\{\hspace{-0.01cm}z\hspace{-0.01cm}\}\hspace{-0.02cm}(\hspace{-0.01cm}t\hspace{-0.01cm}) \hspace{-0.08cm}:=\hspace{-0.1cm}\begin{cases}
\hspace{-0.05cm}\lfloor \hspace{-0.01cm}z(\hspace{-0.01cm}t_0\hspace{-0.01cm})/\tau \hspace{-0.01cm}\rfloor \tau , & \hspace{-0.15cm}\textnormal{if } t_0\hspace{-0.03cm}\leq \hspace{-0.02cm}t\hspace{-0.02cm}<\hspace{-0.03cm}t_1, \\
\hspace{-0.05cm}z(t_l), & \hspace{-0.15cm}\textnormal{if } t_l\hspace{-0.03cm}\leq\hspace{-0.02cm} t\hspace{-0.02cm}<\hspace{-0.03cm}t_{l+1}, l\hspace{-0.04cm}\in\hspace{-0.04cm}\mathbb{N}, 
\end{cases}
\end{equation}
where $\lfloor \cdot \rfloor$ denotes the floor function, and where the sampling points $\{t_l\}_{l=1}^\infty$ are defined by
\begin{equation}
\label{tk}
t_l = \textnormal{inf}\big\{T \in (t_{l-1},\infty)\colon |z(T)-z(t_{l-1})|>\tau\big\}.
\end{equation}
Unlike level-crossing sampling, this setup implies that an output that consecutively crosses the same threshold more than once only triggers one event. This property helps avoid situations where many samples would be sent in a short time-span due to noise. The drawback of this hysteresis feature in the sampling procedure is that it introduces a dynamic nonlinearity; in other words, the event of sending a sample will depend on the previous threshold crossed.

\begin{figure}
	\begin{center}
		\begin{tikzpicture}
		[
		declare function={
			func1(\x)= 	4.5*exp(-\x/2.5)*cos(deg(2*\x));}
		]
		\begin{axis}[
		width=8.2cm, height=5.6cm, axis x line=middle, axis y line=middle,
		ymin=-3, ymax=6, ytick={-2,...,5}, ylabel=$z(t)$,
		xmin=0, xmax=8, xtick={0,...,7}, xlabel=$t${[s]},
		domain=-1:8,samples=301, % added
		]
		\addplot [black,thick,domain=0:7] {func1(x)};
		\addplot [dotted,red] coordinates {(0,4) (0.1616,4)};
		\addplot [dotted,red] coordinates {(0,3.5) (0.2636,3.5)};
		\addplot [dotted,red] coordinates {(0,3) (0.3487,3)};
		\addplot [dotted,red] coordinates {(0,2.5) (0.4258,2.5)};
		\addplot [dotted,red] coordinates {(0,2) (0.4986,2)};
		\addplot [dotted,red] coordinates {(0,1.5) (0.5694,1.5)};
		\addplot [dotted,red] coordinates {(0,1) (3.403,1)};
		\addplot [dotted,red] coordinates {(0,0.5) (3.6748,0.5)};
		\addplot [dotted,red] coordinates {(0,-0.5) (5.0156,-0.5)};
		\addplot [dotted,red] coordinates {(0,-1) (2.0874,-1)};
		\addplot [dotted,red] coordinates {(0,-1.5) (1.9492,-1.5)};
		\addplot [dotted,red] coordinates {(0,-2) (1.7866,-2)};
		\draw[Rect] (0,700) rectangle (16.16,800);
		\draw[Rect] (16.16,650) rectangle (26.36,750);
		\draw[Rect] (26.36,600) rectangle (34.87,700);
		\draw[Rect] (34.87,550) rectangle (42.58,650);
		\draw[Rect] (42.58,500) rectangle (49.86,600);
		\draw[Rect] (49.86,450) rectangle (56.94,550);
		\draw[Rect] (56.94,400) rectangle (63.98,500);
		\draw[Rect] (63.98,350) rectangle (71.13,450);
		\draw[Rect] (71.13,300) rectangle (78.54,400);
		\draw[Rect] (78.54,250) rectangle (86.42,350);
		\draw[Rect] (86.42,200) rectangle (95.09,300);
		\draw[Rect] (95.09,150) rectangle (105.16,250);
		\draw[Rect] (105.16,100) rectangle (118.24,200);
		\draw[Rect] (118.24,50) rectangle (194.92,150);
		\draw[Rect] (194.92,100) rectangle (208.74,200);
		\draw[Rect] (208.74,150) rectangle (221.95,250);
		\draw[Rect] (221.95,200) rectangle (236.6,300);
		\draw[Rect] (236.6,250) rectangle (251,350);
		\draw[Rect] (251,300) rectangle (271.6,400);
		\draw[Rect] (271.6,350) rectangle (367.48,450);
		\draw[Rect] (367.48,300) rectangle (392.7,400);
		\draw[Rect] (392.7,250) rectangle (425.46,350);
		\draw[Rect] (425.46,200) rectangle (549.77,300);
		\draw[Rect] (549.77,250) rectangle (700,350);
		\addplot [only marks,mark=*,red,mark size=1.5] coordinates {(0.1616,4)(0.2636,3.5)(0.3487,3)(0.4258,2.5)(0.4986,2)(0.5694,1.5)(0.6398,1)(2.716,1)(0.7113,0.5)(2.510,0.5)(3.6748,0.5)(0.7854,0)(2.356,0)(3.927,0)(5.4977,0)(0.8642,-0.5)(2.2195,-0.5)(4.2546,-0.5)(0.9509,-1)(2.0874,-1)(1.0516,-1.5)(1.9492,-1.5)(1.1824,-2)};
		\end{axis}
		\end{tikzpicture}
	\end{center}
\vspace{-0.4cm}
	\caption{Lebesgue sampling of a signal $z(t)$ with threshold $\tau = 0.5$. Red dots indicate the sampling instants and thresholds being crossed, and dashed blue rectangles show the regions where $z(t)$ is known to be located.}
	\label{fig3}
\end{figure}
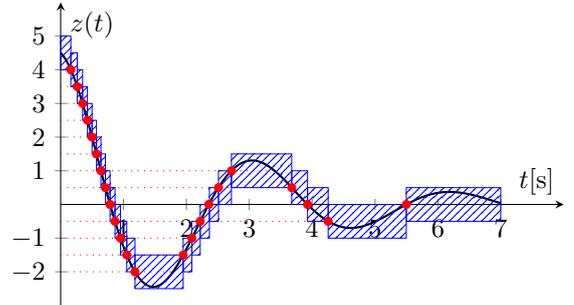

An example of the Lebesgue sampling we consider is shown in Fig. \ref{fig3}. Our goal is to estimate the system parameters that describe the matrices $(\mathbf{A}, \mathbf{B}, \mathbf{C}, D, \mathbf{Q})$ using the input $\{u(t)\}_{t\in [t_1,t_M]}$ and the Lebesgue-sampled output $\{y(t_l)\}_{l=1}^M$. 

\section{Lebesgue-sampled system identification}
\label{sec:lebesguesampledsysid}

In this work, we derive the maximum likelihood estimate of the system matrices using an equivalent discrete-time description of \eqref{system2} for fast sampling rates. The intuition is that the output data $\{y(t_l)\}_{l=1}^M$ provide knowledge of what amplitude band $z(t)$ is located in at any instant of time; thus, the framework is asymptotically equivalent, as the sampling period tends to zero, to computing the maximum likelihood estimate of a system with a fastly-sampled output that is quantized in a special manner. %Thus, we will exploit tools derived from system identification using quantized data~\citep{godoy2011identification}. 

We begin describing our approach by recalling a shift-operator equivalent and an incremental equivalent description of the system \eqref{system2} in Lemma \ref{lemmadiscretization}. In the sequel, we denote any sampled signal $\rho(k\Delta)$ as $\rho_k$, where $k\in \mathbb{N}$. Note that in our application, $k$ ranges between $\lfloor \frac{t_1}{\Delta}\rfloor$ and $\lfloor \frac{t_M}{\Delta}\rfloor$. For simplicity we relabel the discrete-time signal so that $k$ ranges between $k=1$ and $k=N:=\lfloor \frac{t_M}{\Delta}\rfloor-\lfloor \frac{t_1}{\Delta}\rfloor+1$.
\begin{lem}
	\label{lemmadiscretization}
	Consider the continuous-time state-space model in \eqref{system2}, where $u(t)$ is generated by a zero-order-hold with a sampling period $\Delta$. This system has the same second-order output properties at the sampling instants $t=k\Delta$ as the following discrete-time model in shift-operator~form: 
	\begin{subequations}
		\label{ssdq}
		\begin{align}
		\label{ssdq1}
		q\mathbf{x}_k :=\mathbf{x}_{k+1}&= \mathbf{A}_{\textnormal{d}} \mathbf{x}_k + \mathbf{B}_{\textnormal{d}} u_k + \tilde{\mathbf{w}}_k, \\
		\label{ssdq2}
		z_k &= \mathbf{C}\mathbf{x}_k + D u_k,
		\end{align}
	\end{subequations}
	where the matrices $\mathbf{A}_{\textnormal{d}}$ and $\mathbf{B}_{\textnormal{d}}$ are given by
	\begin{equation}
	\mathbf{A}_{\textnormal{d}} = e^{\mathbf{A}\Delta}, \quad \mathbf{B}_{\textnormal{d}} = \int_{0}^\Delta e^{\mathbf{A}s}\textnormal{d}s \mathbf{B},
	\end{equation}
	and the covariance of the noise vector is given by $\mathbb{E}\left\{\tilde{\mathbf{w}}_k {\tilde{\mathbf{w}}_l}^\top\right\} = \mathbf{Q}_{\textnormal{d}} \delta_{k-l}^K$, with
	\begin{equation}
	\label{covariances}
	\mathbf{Q}_{\textnormal{d}} = \int_{0}^\Delta e^{\mathbf{A}s} \mathbf{Q} e^{\mathbf{A}^\top s} \textnormal{d}s.
	\end{equation}
	Alternatively, the discrete-time model derived above can also be written in its incremental or delta-operator form:
	\begin{subequations}
		\label{ssd}
		\begin{align}
		\label{ssd1}
		\textnormal{d}\mathbf{x}_k^+ &= \Delta \mathbf{A}_{\textnormal{in}} \mathbf{x}_k + \Delta \mathbf{B}_{\textnormal{in}} u_k + \textnormal{d}\mathbf{w}_k^+, \\
		\label{ssd2}
		z_k &= \mathbf{C} \mathbf{x}_k + D u_k,
		\end{align}
	\end{subequations}
	where the increments are defined as $\textnormal{d}f_k^+ := f_{k+1} - f_k$, the matrices of the state equation are given by $\mathbf{A}_{\textnormal{in}}=(\mathbf{A}_{\textnormal{d}}-\mathbf{I})/\Delta$, $\mathbf{B}_{\textnormal{in}} = \mathbf{B}_{\textnormal{d}}/\Delta$, and the covariance of the noise vector is $\mathbb{E}\{\textnormal{d}\mathbf{w}_k^+ \textnormal{d}{\mathbf{w}_l^+}^\top\} = \Delta \mathbf{Q}_{\textnormal{in}} \delta_{k-l}^K$, where $\mathbf{Q}_{\textnormal{in}} = \mathbf{Q}_{\textnormal{d}}/\Delta$.
\end{lem}
\begin{pf}
	The proof for the incremental model equivalence can be found in, e.g., \cite[Sec. 3.10]{aastrom1970introduction}, while the equivalent shift-operator model follows from expanding the increment notation in \eqref{ssd} and rearranging terms. \hfill \qed 
\end{pf}
The shift-operator model in \eqref{ssdq} allows for the direct implementation of many filtering and smoothing algorithms. On the other hand, the model in \eqref{ssd} provides a natural way to describe a continuous-time system whose output is sampled at a fast rate, since the continuous-time matrices are recovered when $\Delta \hspace{-0.04cm}\to\hspace{-0.04cm} 0$. That~is,
\begin{equation}
\label{deltatozero}
(\mathbf{A}_{\textnormal{in}},\mathbf{B}_{\textnormal{in}},\mathbf{Q}_{\textnormal{in}})\xrightarrow[]{\Delta \to 0} (\mathbf{A},\mathbf{B},\mathbf{Q}).
\end{equation}

Note that in our approach there is no need to physically include an additional sampling step, and we do not assume that there is a fast-sampling mechanism prior to the event sampler. However, this might be the case in some applications related to incremental encoders, in which $\Delta$ may represent the sampling rate of the high-resolution clock \citep{merry2013optimal}. Also, keep in mind that the sampling points $t_l$ need not be multiples of the fast-sampling period $\Delta$, although this is usually assumed in encoder setups \citep{strijbosch2022iterative}.

The next step is to derive an algorithm that computes~an estimate for the system matrices using the discrete-time equivalent of \eqref{system2}. We use the EM algorithm \citep{dempster1977maximum} for this purpose, which will be applied taking into consideration the shift-operator model~\eqref{ssdq}, as well as the incremental model~\eqref{ssd}, with Lebesgue-sampled output data. To this end, we require formulas for the expectation and maximization steps to implement the EM method. This derivation must include an output perturbation to ensure that the noise covariance matrix in the EM algorithm is full rank \citep{solo2003algorithm}. Thus, we add a zero-mean, independent, Gaussian perturbation $v_k$ with a user-defined variance $\epsilon^2\ll 1$ to $z_k$:
\begin{equation}
\label{zeq}
z_k = \mathbf{C} \mathbf{x}_k + D u_k + v_k.
\end{equation}
We stack the parameters describing the state-space matrices in a vector $\bm{\theta}$, and fix $\hspace{-0.02cm}\{\hspace{-0.02cm}\mathbf{x}_{1:N\hspace{-0.02cm}+\hspace{-0.02cm}1}\hspace{-0.02cm},z_{1:N}\hspace{-0.02cm}\}\hspace{-0.02cm}$ as the set of unobserved latent data. Based on the estimate at the $i$th iteration $\hat{\bm{\theta}}_i$, we need to compute
\begin{equation}
\label{qfunction}
Q(\bm{\theta},\hat{\bm{\theta}}_i) = \mathbb{E}\left\{\log \textnormal{p}(\mathbf{x}_{1:N+1},z_{1:N}|\bm{\theta})|y_{1:N}, \hat{\bm{\theta}}_i \right\},
\end{equation}
where $\textnormal{p}(\mathbf{x}_{1:N+1},z_{1:N}|\bm{\theta})$ is the probability density function (PDF) of $\{\mathbf{x}_{1:N+1},z_{1:N}\}$ given the model parameters $\bm{\theta}$. The function in \eqref{qfunction} is later maximized with respect to $\bm{\theta}$, yielding the new parameter estimate:
\begin{equation}
\hat{\bm{\theta}}_{i+1} = \underset{\bm{\theta}}{\arg \max} \hspace{0.08cm} Q(\bm{\theta},\hat{\bm{\theta}}_i). \notag
\end{equation}
The iterations presented above are performed until the estimate has converged within a predefined tolerance value. In the following subsections, we derive the EM algorithm for the shift and delta operator models and discuss the required filtering and smoothing procedures.

\subsection{E-step for the shift-operator model}
Our goal is to estimate the matrices $\mathbf{A}_{\text{d}},\mathbf{B}_{\text{d}}, \mathbf{C}, D$ and $\mathbf{Q}_\text{d}$, which ultimately must be transformed into their continuous-time equivalents. We assume for this derivation that $\bm{\theta}$ is constituted by the parameters of the shift-operator model matrices. In order to compute the $Q$ function in \eqref{qfunction}, we write the logarithm of the joint probability density function of the extended state vector data $\mathcal{L}(\bm{\theta})=\textnormal{p}( \mathbf{x}_{ 1 : N + 1} , z_{ 1 : N} |\bm{\theta} )$ as
\begin{align*}
- 2 \log \mathcal{L}(\bm{\theta})   &= - 2 \log \textnormal{p}(\mathbf{x}_{ 1} |\bm{\theta} ) - 2 \sum_{k=1}^{N}  \log   \textnormal{p}( \mathbf{x}_{ k + 1} , z_{ k} |\mathbf{x}_{ k} , \bm{\theta} )  \notag  \\
&  =  - 2  \log \left( \frac{\textnormal{p}( \mathbf{x}_{ 1}|\bm{\theta} )}{\sqrt[N/2]{2\pi \epsilon^2}}   \right)  +   N  \log   \det  \left(  2\pi \mathbf{Q}_{\textnormal{d}}   \right)  \\
&+    \frac{1}{\epsilon^2}  \sum_{k=1}^N  \varphi_k^{2}   + \sum_{k=1}^N \mathbf{V}_k^{\top} \mathbf{Q}_{\textnormal{d}}^{-1}\mathbf{V}_k, \notag
\end{align*}
where $\varphi_k\hspace{-0.05cm}=\hspace{-0.05cm}z_k\hspace{-0.05cm}-\hspace{-0.05cm}\mathbf{C}\mathbf{x}_k\hspace{-0.05cm} -\hspace{-0.05cm} D u_k$ and $\mathbf{V}_k\hspace{-0.05cm}=\hspace{-0.05cm}\mathbf{x}_{k+1} \hspace{-0.05cm}-\hspace{-0.05cm}\mathbf{A}_{\textnormal{d}} \mathbf{x}_k \hspace{-0.05cm}-\hspace{-0.05cm} \mathbf{B}_{\textnormal{d}} u_k$. After some embellishments, $Q(\hspace{-0.01cm} \bm{\theta},\hspace{-0.02cm} \hat{\bm{\theta}}_i\hspace{-0.01cm})$ can be expressed as
\begin{align}
&-\hspace{-0.06cm} 2\hspace{-0.02cm}Q(\hspace{-0.01cm} \bm{\theta},\hspace{-0.02cm} \hat{\bm{\theta}}_i\hspace{-0.01cm}) \hspace{-0.08cm}=\hspace{-0.085cm} L_0(\hat{\bm{\theta}}_i) \hspace{-0.03cm}+\hspace{-0.03cm} N\log\det\left(\mathbf{Q}_{\textnormal{d}}\right) \notag \\
&+ \frac{1}{\epsilon^2} \hspace{-0.04cm} \left(\hspace{-0.01cm}D^2 \hspace{-0.01cm}\Gamma_{\hspace{-0.01cm}uu} \hspace{-0.06cm}-\hspace{-0.06cm} 2 \mathbf{C} \mathbf{\Gamma}_{\hspace{-0.01cm}xz}\hspace{-0.06cm}-\hspace{-0.06cm}2 D \Gamma_{\hspace{-0.02cm}uz}\hspace{-0.06cm}+\hspace{-0.06cm}2 D \bm{\Gamma}_{\hspace{-0.01cm}ux}\mathbf{C}^{\hspace{-0.02cm}\top}\hspace{-0.06cm}+\hspace{-0.06cm}\mathbf{C}\bm{\Gamma}_{\hspace{-0.02cm}xx}\mathbf{C}^{\hspace{-0.02cm}\top} \hspace{-0.01cm}\right) \notag \\
&+\hspace{-0.02cm}\textnormal{tr} \bigg\{ \hspace{-0.05cm}\mathbf{Q}_{\textnormal{d}}^{-\hspace{-0.02cm}1}\hspace{-0.04cm} \bigg(\hspace{-0.06cm}\bm{\Gamma}_{qq} \hspace{-0.07cm}+\hspace{-0.07cm} \mathbf{A}_{\textnormal{d}} \bm{\Gamma}_{\hspace{-0.02cm}xx} \mathbf{A}_{\textnormal{d}}^\top \hspace{-0.07cm}+\hspace{-0.07cm} \mathbf{B}_{\textnormal{d}} \Gamma_{\hspace{-0.02cm}uu} \mathbf{B}_{\textnormal{d}}^\top \hspace{-0.07cm}-\hspace{-0.07cm} \mathbf{A}_{\textnormal{d}}\bm{\Gamma}_{\hspace{-0.02cm}x q}\hspace{-0.07cm}-\hspace{-0.07cm}\bm{\Gamma}_{\hspace{-0.02cm}xq}^\top \mathbf{A}_{\textnormal{d}}^\top  \notag \\
&\hspace{0.8cm}- \mathbf{B}_{\textnormal{d}} \bm{\Gamma}_{u q} - \bm{\Gamma}_{uq}^\top \mathbf{B}_{\textnormal{d}}^\top + \mathbf{A}_{\textnormal{d}} \bm{\Gamma}_{ux}^\top\mathbf{B}_\textnormal{d}^\top + \mathbf{B}_\textnormal{d} \bm{\Gamma}_{ux} \mathbf{A}_\textnormal{d}^\top \bigg)\bigg\}, \notag 
\end{align}
where $L_0(\hat{\bm{\theta}}_i)$ accounts for all terms solely depending on $\hat{\bm{\theta}}_i$ or constants, and
\begin{subequations}
	\label{filtering}
	\begin{align}
	\label{filtering1}
	\bm{\Gamma}_{xx} &=  \bar{\mathbb{E}}\{\mathbf{x}_k\mathbf{x}_k^{\top} \}, 	&&\bm{\Gamma}_{qq} = \bar{\mathbb{E}}\{\mathbf{x}_{k+1}\mathbf{x}_{k+1}^\top \},\\
	\label{filtering2}
	\bm{\Gamma}_{x q}  &= \bar{\mathbb{E}}\{\mathbf{x}_k\mathbf{x}_{k+1}^\top  \},
	&&\bm{\Gamma}_{u x}  = \bar{\mathbb{E}}\{u_k\mathbf{x}_k^\top \}, \\
	\label{filtering3}
	\bm{\Gamma}_{u q}  &= \bar{\mathbb{E}}\{u_k\mathbf{x}_{k+1}^\top \},
	&&\bm{\Gamma}_{xz}  = \bar{\mathbb{E}}\{\mathbf{x}_k z_k \}, \\
	\label{filtering4}
	\Gamma_{uz}  &=  \bar{\mathbb{E}}\{u_k z_k \}, &&\Gamma_{uu}  = \bar{\mathbb{E}}\{u_{k}^2\},
	\end{align}
\end{subequations}
where $\bar{\mathbb{E}}\{\cdot\}:=\sum_{k=1}^{N}\mathbb{E}\{\cdot|y_{1:N}, \hat{\bm{\theta}}_i \}$.
\subsection{Computation of the state moments} 
	In order to compute the quantities in \eqref{filtering} that determine the auxiliary function $Q$, it is necessary to compute some moments related to the system state $\mathbf{x}_k$, the output $z_k$, and their corresponding cross moments. This computation requires the evaluation of filtering and smoothing distributions of the extended vector $\mathbf{x}_{k}^{\textrm{e}}:=[ \begin{matrix} \mathbf{x}_{k}^\top & z_{k}\end{matrix}]^\top$ conditioned on the measured data, i.e., the PDFs $\textnormal{p}(\mathbf{x}_{k}^{\textrm{e}}|y_{1:k})$ and $\textnormal{p}(\mathbf{x}_{k}^{\textrm{e}}|y_{1:N})$. Several methods allow the computation of these PDFs; one recent and promising approach is the Gaussian Sum Filter and Smoother developed in \cite{Cedeno2021b,Cedeno2021a}, where the desired PDFs are represented by a Gaussian Mixture Model. An alternative approach to obtain the moments of $\mathbf{x}_k$ and $z_k$ is the Sequential Monte Carlo sampling approach, also called particle filter/smoother (PF/PS) \citep{Gordon1993,Doucet2000}. In this approach, filtering and smoothing distributions are represented by using a set of weighted random samples called particles so that
	\begin{align}
		&\textnormal{p}(\mathbf{x}_{k}|y_{1:k}) \approx \sum_{i=1}^{\mathcal{M}} w_k^{(i)} \delta\left(\mathbf{x}_k-\mathbf{x}_k^{(i)}\right), \\
        &\textnormal{p}(\mathbf{x}_{k}|y_{1:N}) \approx \sum_{i=1}^{\mathcal{M}} w_{k|N}^{(i)} \delta\left( \mathbf{x}_k-\tilde{\mathbf{x}}_k^{(i)}\right),
	\end{align}
	where $w_k^{(i)}$ and $w_{k|N}^{(i)}$ denote the $i$th weights, and $\mathbf{x}_k^{(i)}$ with $\tilde{\mathbf{x}}_k^{(i)}$ denote the $i$th particles sampled from the filtering and smoothing PDFs $\textnormal{p}(\mathbf{x}_{k}|y_{1:k})$ and $\textnormal{p}(\mathbf{x}_{k}|y_{1:N})$, respectively. The quantity $\mathcal{M}$ is the number of particles.  The importance weight computation can be carried out recursively (Sequential Importance Sampling) as follows:
	\begin{align}\label{eqn:filter_weights}
		w_k^{(i)}  &\propto w_{k-1}^{(i)} \dfrac{\textnormal{p}(y_k|\mathbf{x}_t^{(i)})\textnormal{p}(\mathbf{x}_k^{(i)}|\mathbf{x}_{k-1}^{(i)})}{\textnormal{h}(\mathbf{x}_k|\mathbf{x}^{(i)}_{k-1},y_k)}, \notag \\
		w_{k|N}^{(i)} &= \sum_{j=1}^{\mathcal{M}} w_{k+1|N}^{(j)}\dfrac{w_{k}^{(i)}\textnormal{p}(\mathbf{x}_{k+1}^{(j)}|\mathbf{x}_{k}^{(i)})}{\sum_{k=1}^{\mathcal{M}}w_{k}^{(k)}\textnormal{p}(\mathbf{x}_{k+1}^{(j)}|\mathbf{x}_{k}^{(k)})}, \notag 
	\end{align}
	where $\textnormal{h}(\mathbf{x}_k|\mathbf{x}_{k-1}^{(i)},y_k)$ is the importance density, $w_{N|N}^{(i)}=w_{N}^{(i)}$ for $i=1,\dots,\mathcal{M}$, and $w_{k-1}^{(i)}$ are the importance weights of the previous iteration. Details of the implementation and comparison of particle filters/smoothers to deal with quantized data can be found in \cite{Cedeno2023}. Once the PF and PS are implemented, the particles of the smoothing distributions can be used to approximate the moments and cross moments of $\mathbf{x}_k$ and $z_k$ as follows: 
	\begin{equation}
		\mathbb{E}\left\lbrace g(\mathbf{x}_k)|y_{1:N}\right\rbrace \approx \sum_{i=1}^{\mathcal{M}}w_{k|N}^{(i)}g(\tilde{\mathbf{x}}_k^{(i)}), \notag
	\end{equation}
	where $w_{k|N}^{(i)}$ and $\tilde{\mathbf{x}}_k^{(i)}$ are the weights and particles (from the PS), and $g(\mathbf{x}_k)$ is a function of $\mathbf{x}_k$, for instance $g(\mathbf{x}_k)=\mathbf{x}_k$ or  $g(\mathbf{x}_k)=\left(\mathbf{x}_k-\mathbb{E}\left\lbrace \mathbf{x}_k\right\rbrace\right)\left(\mathbf{x}_k-\mathbb{E}\left\lbrace \mathbf{x}_k\right\rbrace\right)^{\hspace{-0.04cm}\top}$. The corresponding moments of $z_k$ are approximated by
	\begin{equation}
	    \mathbb{E}\{\hspace{-0.02cm}z_k|y_{1\hspace{-0.01cm}:\hspace{-0.01cm}N}\hspace{-0.02cm}\} \hspace{-0.09cm}\approx\hspace{-0.07cm} \sum_{i=1}^{\mathcal{M}}\hspace{-0.07cm}w_{k|\hspace{-0.02cm}N}^{(i)}\hat{z}_k, \hspace{0.2cm} 
		\mathbb{E} \{\hspace{-0.02cm}\mathbf{x}_k \hspace{-0.02cm}z_k|y_{1\hspace{-0.01cm}:\hspace{-0.01cm}N}\hspace{-0.02cm}\} \hspace{-0.09cm}\approx \hspace{-0.07cm} \sum_{i=1}^{\mathcal{M}}w_{k|N}^{(i)}\tilde{\mathbf{x}}_k^{(i)}\hat{z}_k, \notag
	\end{equation}
	with $\hat{z}_k=\mathbb{E}\{z_k|\tilde{\mathbf{x}}_k^{(i)},y_{k}\}$ being the mean of the truncated Gaussian distribution given by
	\begin{align}
		\hat{z}_k &= \mathbf{C}\tilde{\mathbf{x}}_k^{(i)}+Du_k+\epsilon \dfrac{\psi(a_k,b_k)}{\Psi(a_k,b_k)},  \notag \\
		\psi(a_k,b_k)\hspace{-0.06cm}&=\hspace{-0.06cm}\phi\left[\hspace{-0.05cm}\frac{a_k\hspace{-0.05cm}-\hspace{-0.05cm}\mathbf{C}\tilde{\mathbf{x}}_k^{(i)}\hspace{-0.05cm}-\hspace{-0.05cm}Du_k}{\epsilon}\hspace{-0.05cm}\right]\hspace{-0.05cm}-\hspace{-0.05cm}\phi\left[\hspace{-0.05cm}\frac{b_k\hspace{-0.05cm}-\hspace{-0.05cm}\mathbf{C}\tilde{\mathbf{x}}_k^{(i)}\hspace{-0.05cm}-\hspace{-0.05cm}Du_k}{\epsilon}\hspace{-0.05cm}\right], \notag
	\end{align}
	\begin{align}
		\Psi(a_k,b_k)\hspace{-0.06cm}&=\hspace{-0.06cm}\Phi\left[\hspace{-0.05cm}\frac{a_k\hspace{-0.05cm}-\hspace{-0.05cm}\mathbf{C}\tilde{\mathbf{x}}_k^{(i)}\hspace{-0.05cm}-\hspace{-0.05cm}Du_k}{\epsilon}\hspace{-0.05cm}\right]\hspace{-0.05cm}-\hspace{-0.05cm}\Phi\left[\hspace{-0.05cm}\frac{b_k\hspace{-0.05cm}-\hspace{-0.05cm}\mathbf{C}\tilde{\mathbf{x}}_k^{(i)}\hspace{-0.05cm}-\hspace{-0.05cm}Du_k}{\epsilon}\hspace{-0.05cm}\right]\hspace{-0.03cm}, \notag
	\end{align}
	where $[a_k,b_k]$ is an interval defined by the threshold regions and the output $y_k$, $\phi[x]$ is the standard normal density, and $\Phi[x]$ is its cumulative distribution function.

\subsection{M-step for the shift-operator model}
Now we need to maximize $Q(\bm{\theta},\hat{\bm{\theta}}_i)$ with respect to $\bm{\theta}$. The following result provides the EM iterations that are proposed for estimating the discrete-time equivalent of the state-space matrices of interest.

\begin{thm}
	\label{lemmaEM}
	The matrices that maximize $Q(\bm{\theta},\hat{\bm{\theta}}_i)$ for the shift-operator model are given by
	\begin{align}\label{em1}
	&\begin{bmatrix}
	\mathbf{A}_{\textnormal{d},i+1} & \mathbf{B}_{\textnormal{d},i+1} \\
	\mathbf{C}_{i+1} & D_{i+1} \\
	\end{bmatrix} \!=\! \begin{bmatrix}
	\bm{\Gamma}_{x q} & \bm{\Gamma}_{x z} \\
	\bm{\Gamma}_{u q} & \Gamma_{u z}
	\end{bmatrix}^\top 
	\begin{bmatrix}
	\bm{\Gamma}_{xx} &  \bm{\Gamma}_{ux}^\top \\
	\bm{\Gamma}_{ux} &  \Gamma_{uu}
	\end{bmatrix}^{-1}, \\
	\label{em2}
	&\mathbf{Q}_{\textnormal{d},i+1} \!= \! \frac{1}{N}  \left(\bm{\Gamma}_{ qq} - \begin{bmatrix}
	\bm{\Gamma}_{x q} \\ \bm{\Gamma}_{uq} 
	\end{bmatrix}^{\top} 
	\begin{bmatrix}
	\bm{\Gamma}_{xx} & \bm{\Gamma}_{ux}^\top \\
	\bm{\Gamma}_{ux} & \Gamma_{uu}
	\end{bmatrix}^{-1} \begin{bmatrix}
	\bm{\Gamma}_{x q} \\ \bm{\Gamma}_{u q}
	\end{bmatrix} \right),
	\end{align}	
	where all the $\bm{\Gamma}$ matrices are given in \eqref{filtering}.
\end{thm}
\begin{pf}
See Appendix A. \hspace{4.2cm} \hfill \qed 
\end{pf}

\subsection{EM algorithm for the incremental model}

Theorem \ref{lemmaEM} provides the EM iterations for the shift-operator model \eqref{ssdq}. However, since our interest is in the continuous-time state-space matrices, we must revert the discretization process of Lemma \ref{lemmadiscretization} in some way. One alternative is to estimate the discrete-time matrices and later transform them to continuous-time, although this step is known to be ill-conditioned for small sampling periods \citep{garnier2014advantages}. Another option for small $\Delta$ is to derive the EM iterations for the incremental model instead, and let the incremental state-space matrices represent the continuous-time ones according to \eqref{deltatozero}, i.e., $(\mathbf{A}_{\textnormal{in}},\mathbf{B}_{\textnormal{in}},\mathbf{Q}_{\textnormal{in}})\approx (\mathbf{A},\mathbf{B},\mathbf{Q})$. This is the approach suggested by \cite{yuz2011identification} for the identification of stochastic differential equations, and it is convenient since it leads to explicit iterations of the EM algorithm for the continuous-time system parameters, as seen next.

\begin{thm}
	\label{theoremEM}    
	The EM iterations for the incremental model \eqref{ssd} with Lebesgue-sampled output data are given by
	\begin{align}
	\label{emdelta1}
	&\begin{bmatrix}
	\mathbf{A}_{\textnormal{in},i+1} &  \mathbf{B}_{\textnormal{in},i+1} \\
	\mathbf{C}_{i+1} &  D_{i+1} \\
	\end{bmatrix}  \!= \! \begin{bmatrix}
	\bm{\Gamma}_{x \delta} &  \bm{\Gamma}_{x z} \\
	\bm{\Gamma}_{u \delta} &  \Gamma_{u z}
	\end{bmatrix}^\top 
	\begin{bmatrix}
	\bm{\Gamma}_{xx} & \bm{\Gamma}_{ux}^\top \\
	\bm{\Gamma}_{ux} & \Gamma_{uu}
	\end{bmatrix}^{-1}, \\
	\label{emdelta2}
	&\mathbf{Q}_{\textnormal{in},i+1} \!= \!\frac{\Delta}{N} \left(\bm{\Gamma}_{\delta \delta} - \begin{bmatrix}
	\bm{\Gamma}_{x \delta} \\ \bm{\Gamma}_{u\delta} 
	\end{bmatrix}^\top 
	\begin{bmatrix}
	\bm{\Gamma}_{xx} & \bm{\Gamma}_{ux}^\top \\
	\bm{\Gamma}_{ux} &  \Gamma_{uu}
	\end{bmatrix}^{-1}  \begin{bmatrix}
	\bm{\Gamma}_{x \delta} \\ \bm{\Gamma}_{u \delta}
	\end{bmatrix} \right),
	\end{align}
	where $\bm{\Gamma}_{xx}, \bm{\Gamma}_{ux}, \bm{\Gamma}_{xz}, \Gamma_{uz}$  and $\Gamma_{uu}$ are given in \eqref{filtering}, and
	\begin{subequations}
		\label{deltafiltering}
		\begin{align}
		\bm{\Gamma}_{x \delta} &= \Delta^{-1}\bar{\mathbb{E}}\left\{\mathbf{x}_k(\mathbf{x}_{k+1}-\mathbf{x}_k)^\top \right\}, \\
		\bm{\Gamma}_{u\delta}  &=  \Delta^{-1} \bar{\mathbb{E}}\left\{u_k(\mathbf{x}_{k+1}-\mathbf{x}_k)^\top \right\}, \\
		\bm{\Gamma}_{\delta \delta}  &=  \Delta^{-2} \bar{\mathbb{E}}\left\{(\mathbf{x}_{k+1}-\mathbf{x}_k)(\mathbf{x}_{k+1}-\mathbf{x}_k)^{\top} \right\}.
		\end{align}
	\end{subequations}
\end{thm}
\begin{pf}
See Appendix B. \hspace{4.2cm} \hfill \qed
\end{pf}

It is well known that the converging point of EM iterations belongs to the set of stationary points of the likelihood function. Thus, if adequately initialized and under mild conditions regarding the model structure and input excitation, the iterations in Theorem \ref{theoremEM} provide the maximum likelihood estimate of the system parameters in its incremental form. These estimates are biased estimates of the continuous-time parameters due to the approximation $(\mathbf{A}_{\textnormal{in}},\mathbf{B}_{\textnormal{in}},\mathbf{Q}_{\textnormal{in}})\approx (\mathbf{A},\mathbf{B},\mathbf{Q})$. This bias can be shown to be proportional to the fast-sampling period $\Delta$; explicit bounds will be published elsewhere.

\section{Simulations}
\label{sec:simulations}

\begin{figure*}
	\centering
	\includegraphics[width=1\linewidth]{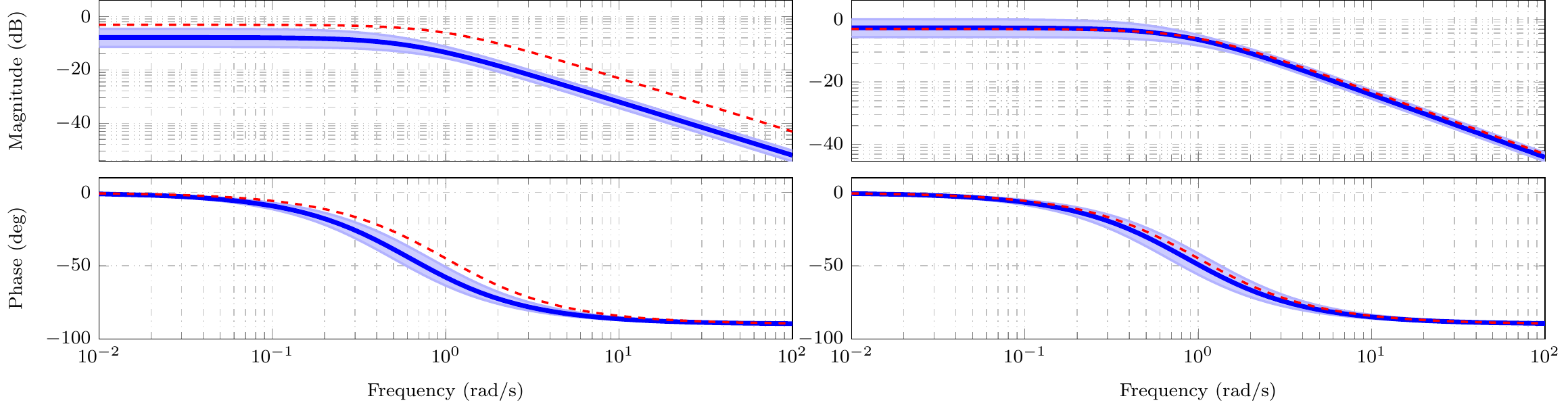}
	\vspace{-0.5cm}
	\caption{Frequency response of the true and estimated system. Left: KS-EM; right: PS-EM (proposed method).}
	\label{fig:bode_escalar_n2000}
\end{figure*}

\begin{figure}
	\centering
	\includegraphics[width=0.89\linewidth]{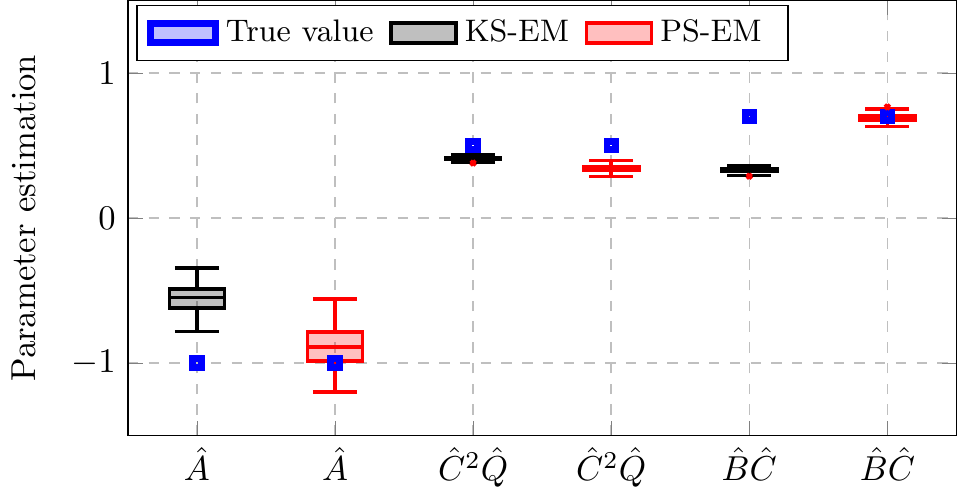}
	\vspace{-0.1cm}
	\caption{Estimates of the system parameters that are invariant under similarity transformations.}
	\label{fig:escalar_boxplots_N2000}
\end{figure}

In this section, we present a numerical example to analyze the performance of the proposed method (PS-EM), in which we utilize the particle smoother to compute the EM iterations. We compare PS-EM to the standard method to identify state-space models with the Kalman Smoother (KS-EM, \cite{gibson2005robust}), which does not consider any type of quantization. The continuous-time system we consider is given by
	\begin{equation}
	\dot{x}(t) = -x(t) + 0.7u(t) + \dot{w}(t), \hspace{0.3cm} z(t) = x(t), \notag
	\end{equation}
where the measured output is computed from \eqref{Qh} with $\tau=0.3$, $\dot{w}(t)\sim \mathcal{N}(\dot{w}(t);0,0.5)$, and the input is sampled from $\mathcal{N}(u(t);0,\sigma^2)$ with $\sigma=10$. We use $N=2000$ and $\Delta=0.01$, and to compute the moments of $x_k$ and $z_k$, we implement the particle filter with $\mathcal{M}=1000$ particles. The frequency response of the true system and 100 Monte Carlo runs are shown in Fig. \ref{fig:bode_escalar_n2000}. The blue-shaded region represents the area where all Monte Carlo frequency responses lie, and the left and right plots show the response obtained with KS-EM and PS-EM, respectively. The results show that our approach produces more accurate estimates of the system than the KS-EM method. In addition, Fig. \ref{fig:escalar_boxplots_N2000} presents boxplots of the parameter estimates that are invariant under similarity transformations. The proposed method PS-EM outperforms KS-EM in the system parameter estimation and is competitive against KS-EM in estimating the noise covariance matrix. Systems of higher order have also been tested and show similar results to this case study. In terms of computation time, the burden of our approach using $1000$ particles and $2000$ data points was approximately $36$ minutes per Monte Carlo run. This agrees with the execution time reported for particle smoothing in, e.g., \cite{Cedeno2021b,Cedeno2023} multiplied by the total number of the EM iterations. The computer used has an Intel(R) Core(TM) i5-8300H CPU @ 2.30 GHz processor, and a RAM of 8.00 GB, with Windows 11 and MATLAB 2021b.


\section{Conclusions}
\label{sec:conclusions}

This paper introduces an identification method for continuous-time LTI systems with Lebesgue-sampled observations based on the maximum likelihood principle. We have used the EM algorithm to derive an iterative procedure that obtains an estimate of the system parameters and the noise covariance matrix. For this, we implemented a particle filter that evaluates the filtering and smoothing distributions that are required to compute the auxiliary function of the EM algorithm. The proposed method was compared to the standard approach that does not take into account the Lebesgue sampling of the output data, verifying that the former approach yields more accurate estimates of the system than the latter approach.

\section*{Appendix}
\subsection{Proof of Theorem \ref{lemmaEM}}
\label{appendixA}

	We first derive \eqref{em1}. If we consider only the terms that are dependent on $\mathbf{A}_{\text{d}}$ and $\mathbf{B}_\text{d}$, $Q(\bm{\theta},\hat{\bm{\theta}}_i)$ is proportional to
	\begin{align}
	-Q(\bm{\theta},\hat{\bm{\theta}}_i) &\propto \textnormal{tr} \Bigg\{\mathbf{Q}_\text{d}^{-1} \bigg(\begin{bmatrix}
	\mathbf{A}_\text{d} & \hspace{-0.05cm}\mathbf{B}_\text{d}
	\end{bmatrix}  \hspace{-0.04cm}
	\begin{bmatrix}
	\bm{\Gamma}_{xx} & \bm{\Gamma}_{ux}^\top \\
	\bm{\Gamma}_{ux} & \Gamma_{uu}
	\end{bmatrix}  \hspace{-0.04cm}
	\begin{bmatrix}
	\mathbf{A}_\text{d}^\top \\
	\mathbf{B}_\text{d}^\top
	\end{bmatrix} \notag \\
	\label{minQsystem}
	&- \hspace{-0.05cm} 
	\begin{bmatrix}
	\mathbf{A}_\text{d} & \hspace{-0.05cm}\mathbf{B}_\text{d}
	\end{bmatrix} \hspace{-0.05cm}
	\begin{bmatrix}
	\bm{\Gamma}_{x q} \\
	\bm{\Gamma}_{u q}
	\end{bmatrix} \hspace{-0.07cm}
	-\hspace{-0.06cm} \begin{bmatrix}
	\bm{\Gamma}_{xq}^\top & \hspace{-0.05cm} \bm{\Gamma}_{uq}^\top
	\end{bmatrix} \hspace{-0.07cm}
	\begin{bmatrix}
	\mathbf{A}_\text{d}^\top \\
	\mathbf{B}_\text{d}^\top
	\end{bmatrix}\hspace{-0.03cm} \bigg) \hspace{-0.05cm}\Bigg\}.
	\end{align}
	Using the fact that for any positive definite matrix $\mathbf{A}$ and any matrices $\mathbf{B}$ and $\mathbf{X}$ of suitable dimension we have
	\begin{align}
	\mathbf{X}\mathbf{A}\mathbf{X}^{\hspace{-0.05cm}\top} \hspace{-0.19cm} - \hspace{-0.1cm} \mathbf{X}\mathbf{B} \hspace{-0.09cm}-\hspace{-0.09cm} \mathbf{B}^{\hspace{-0.05cm}\top}\hspace{-0.06cm} \mathbf{X}^{\hspace{-0.05cm}\top} \hspace{-0.16cm}&=\hspace{-0.09cm} (\hspace{-0.015cm}\mathbf{X}\hspace{-0.11cm}-\hspace{-0.08cm}\mathbf{B}^{\hspace{-0.05cm}\top} \hspace{-0.1cm}\mathbf{A}^{\hspace{-0.05cm}-\hspace{-0.02cm}1}\hspace{-0.035cm}) \hspace{-0.01cm}\mathbf{A}\hspace{-0.01cm}(\hspace{-0.015cm}\mathbf{X}\hspace{-0.11cm}-\hspace{-0.08cm}\mathbf{B}^{\hspace{-0.05cm}\top} \hspace{-0.1cm}\mathbf{A}^{\hspace{-0.05cm}-\hspace{-0.02cm}1}\hspace{-0.035cm}) ^{\hspace{-0.05cm}\top} \hspace{-0.16cm}- \hspace{-0.09cm}\mathbf{B}^{\hspace{-0.05cm}\top} \hspace{-0.09cm}\mathbf{A}^{\hspace{-0.06cm}-\hspace{-0.02cm}1}\hspace{-0.03cm}\mathbf{B} \notag \\
	&\succeq -\mathbf{B}^\top \mathbf{A}^{-1}\mathbf{B}, \notag
	\end{align}
	we exploit the partial ordering of the trace \cite[Corollary 7.7.4]{Horn2012} to conclude that the minimum of \eqref{minQsystem} is achieved when $\mathbf{A}_{\textnormal{d}}$ and $\mathbf{B}_{\textnormal{d}}$ are as in \eqref{em1}.
	
	Proceeding similarly with $\mathbf{C}$ and $D$, we find that
	\small
	\begin{equation}
	-Q(\bm{\theta},\hspace{-0.02cm}\hat{\bm{\theta}}_i) \hspace{-0.04cm}\propto \hspace{-0.06cm} 
	\begin{bmatrix}
	\mathbf{C}^{\hspace{-0.03cm}\top} \\ D
	\end{bmatrix}^{\hspace{-0.06cm}\top}  \hspace{-0.12cm}
	\begin{bmatrix}
	\bm{\Gamma}_{\hspace{-0.03cm}xx} & \hspace{-0.05cm}\bm{\Gamma}_{\hspace{-0.03cm}ux}^\top \\
	\bm{\Gamma}_{\hspace{-0.03cm}ux} & \hspace{-0.05cm}\Gamma_{\hspace{-0.03cm}uu}
	\end{bmatrix}  \hspace{-0.07cm}
	\begin{bmatrix}
	\mathbf{C}^{\hspace{-0.03cm}\top} \\ D
	\end{bmatrix} \hspace{-0.05cm} - \hspace{-0.05cm} 
	\begin{bmatrix}
	\mathbf{C}^{\hspace{-0.03cm}\top} \\ D
	\end{bmatrix}^{\hspace{-0.06cm}\top}  \hspace{-0.12cm}
	\begin{bmatrix}
	\bm{\Gamma}_{\hspace{-0.03cm}x z} \\
	\bm{\Gamma}_{\hspace{-0.03cm}u z}
	\end{bmatrix} \hspace{-0.05cm}
	-\hspace{-0.05cm} \begin{bmatrix}
	\bm{\Gamma}_{\hspace{-0.03cm}xz} \\ \Gamma_{\hspace{-0.03cm}uz}
	\end{bmatrix}^{\hspace{-0.06cm}\top} \hspace{-0.12cm}
	\begin{bmatrix}
	\mathbf{C}^\top \\
	D
	\end{bmatrix}\hspace{-0.08cm}. \notag
	\end{equation}
	\normalsize
	By the same reasoning as above, the minimum is achieved when $\mathbf{C}$ and $D$ are as in \eqref{em1}. To obtain \eqref{em2}, we note that when $\mathbf{A}_{\textnormal{d},i+1}$ and $\mathbf{B}_{\textnormal{d},i+1}$ are inserted in $Q(\bm{\theta},\hat{\bm{\theta}}_i)$ we have
	\begin{align}
	&-Q(\bm{\theta},\hat{\bm{\theta}}_i) \propto N\log \det(\mathbf{Q}_\textnormal{d})\notag \\
	\label{expressionabove}
	&\hspace{0.1cm}+\textnormal{tr}\left\{\mathbf{Q}_\textnormal{d}^{-1} \hspace{-0.1cm}\left(\hspace{-0.03cm}\bm{\Gamma}_{q q}\hspace{-0.03cm} -\hspace{-0.03cm} \begin{bmatrix}
	\bm{\Gamma}_{x q} \\ \bm{\Gamma}_{uq} 
	\end{bmatrix}^\top \hspace{-0.05cm}
	\begin{bmatrix}
	\bm{\Gamma}_{xx} & \hspace{-0.05cm}\bm{\Gamma}_{ux}^\top \\
	\bm{\Gamma}_{ux} & \hspace{-0.05cm} \Gamma_{uu}
	\end{bmatrix}^{\hspace{-0.03cm}-\hspace{-0.02cm}1} \hspace{-0.05cm} \begin{bmatrix}
	\bm{\Gamma}_{x q} \\ \bm{\Gamma}_{u q}
	\end{bmatrix} \hspace{-0.03cm}\right) \right\}\hspace{-0.03cm}. \hspace{-0.1cm}
	\end{align}
	Setting to zero the partial derivative of \eqref{expressionabove} with respect to $\mathbf{Q}_\textnormal{d}$ indicates that $\mathbf{Q}_{\textnormal{d},i+1}$ must satisfy
	\begin{equation}
	\mathbf{Q}_{\textnormal{d},\hspace{-0.02cm}i\hspace{-0.02cm}+\hspace{-0.02cm}1}^{-1}\hspace{-0.12cm}\left(\hspace{-0.11cm}N\hspace{-0.02cm}\mathbf{Q}_{\textnormal{d},\hspace{-0.02cm}i\hspace{-0.02cm}+\hspace{-0.02cm}1}\hspace{-0.11cm}-\hspace{-0.08cm}\bm{\Gamma}_{\hspace{-0.04cm}xx}\hspace{-0.1cm}+\hspace{-0.09cm}\begin{bmatrix}
	\bm{\Gamma}_{\hspace{-0.04cm}xq} \\ \bm{\Gamma}_{\hspace{-0.04cm}uq} 
	\end{bmatrix}^{\hspace{-0.07cm}\top}\hspace{-0.14cm}
	\begin{bmatrix}
	\bm{\Gamma}_{\hspace{-0.04cm}xx} & \hspace{-0.05cm}\bm{\Gamma}_{\hspace{-0.04cm}ux}^\top \\
	\bm{\Gamma}_{\hspace{-0.04cm}ux} & \hspace{-0.05cm} \Gamma_{\hspace{-0.04cm}uu}
	\end{bmatrix}^{\hspace{-0.06cm}-\hspace{-0.02cm}1} \hspace{-0.12cm}\begin{bmatrix}
	\bm{\Gamma}_{\hspace{-0.04cm}xq} \\ \bm{\Gamma}_{\hspace{-0.04cm}uq} 
	\end{bmatrix} \hspace{-0.03cm}\right) \hspace{-0.1cm}\mathbf{Q}_{\textnormal{d},\hspace{-0.02cm}i\hspace{-0.02cm}+\hspace{-0.02cm}1}^{-1} \hspace{-0.13cm}=\hspace{-0.07cm} \mathbf{0}, \notag
	\end{equation}
	which directly leads to \eqref{em2}, concluding the proof. \hspace{0.4cm}\hfill\qed

\subsection{Proof of Theorem \ref{theoremEM}}
\label{prooftheoremem}
	By leveraging the fact that $\mathbf{A}_{\textnormal{d}}=\mathbf{I}+\Delta\mathbf{A}_{\textnormal{in}}$ and $\mathbf{B}_{\textnormal{d}} = \Delta\mathbf{B}_{\textnormal{in}}$, the first block row of \eqref{em1} can be expressed as
	\begin{align}
	&\Delta\hspace{-0.04cm} \begin{bmatrix}
	\mathbf{A}_{\textnormal{in},i\hspace{-0.02cm}+\hspace{-0.02cm}1} &  \hspace{-0.05cm}\mathbf{B}_{\textnormal{in},i\hspace{-0.02cm}+\hspace{-0.02cm}1}
	\end{bmatrix} \hspace{-0.09cm}+\hspace{-0.07cm}\begin{bmatrix}
	\mathbf{I} & \hspace{0.05cm}\mathbf{0}
	\end{bmatrix} \hspace{-0.09cm}= \hspace{-0.07cm}\begin{bmatrix}
	\bm{\Gamma}_{\hspace{-0.04cm}xq} \\ \bm{\Gamma}_{\hspace{-0.04cm}uq} 
	\end{bmatrix}^{\hspace{-0.06cm}\top}\hspace{-0.12cm}
	\begin{bmatrix}
	\bm{\Gamma}_{\hspace{-0.04cm}xx} & \hspace{-0.05cm}\bm{\Gamma}_{\hspace{-0.04cm}ux}^\top \\
	\bm{\Gamma}_{\hspace{-0.04cm}ux} & \hspace{-0.05cm} \Gamma_{\hspace{-0.04cm}uu}
	\end{bmatrix}^{\hspace{-0.04cm}-\hspace{-0.02cm}1} \notag \\
	&\implies \hspace{-0.12cm}\begin{bmatrix}
	\mathbf{A}_{\textnormal{in},i\hspace{-0.02cm}+\hspace{-0.02cm}1} & \hspace{-0.05cm}\mathbf{B}_{\textnormal{in},i\hspace{-0.02cm}+\hspace{-0.02cm}1}
	\end{bmatrix} \hspace{-0.1cm} = \hspace{-0.08cm} \frac{1}{\Delta} \hspace{-0.08cm}\left(\hspace{-0.06cm}\begin{bmatrix}
	\bm{\Gamma}_{\hspace{-0.03cm}x q} \\
	\bm{\Gamma}_{\hspace{-0.03cm}u q} 
	\end{bmatrix}\hspace{-0.08cm}-\hspace{-0.08cm}\begin{bmatrix}
	\bm{\Gamma}_{\hspace{-0.03cm}x x} \\
	\bm{\Gamma}_{\hspace{-0.03cm}u x} 
	\end{bmatrix}\hspace{-0.05cm}\right)^{\hspace{-0.09cm}\top} 
	\hspace{-0.09cm}
	\begin{bmatrix}
	\bm{\Gamma}_{\hspace{-0.03cm}xx} & \hspace{-0.05cm}\bm{\Gamma}_{\hspace{-0.03cm}ux}^\top \\
	\bm{\Gamma}_{\hspace{-0.03cm}ux} & \hspace{-0.05cm}\Gamma_{\hspace{-0.03cm}uu}
	\end{bmatrix}^{\hspace{-0.03cm}-\hspace{-0.02cm}1} \notag \\
	&\hspace{3.12cm}= \begin{bmatrix}
	\bm{\Gamma}_{x \delta} \\
	\bm{\Gamma}_{u \delta} 
	\end{bmatrix}^\top 
	\begin{bmatrix}
	\bm{\Gamma}_{xx} & \hspace{-0.05cm}\bm{\Gamma}_{ux}^\top \\
	\bm{\Gamma}_{ux} & \hspace{-0.05cm}\Gamma_{uu}
	\end{bmatrix}^{-1}, \notag
	\end{align}
	where $\bm{\Gamma}_{x \delta}$ and $\bm{\Gamma}_{u \delta}$ are as in \eqref{deltafiltering}. Similarly, we can write $\mathbf{Q}_{\textnormal{d},i+1}$ in terms of $\bm{\Gamma}_{\delta \delta}, \bm{\Gamma}_{x \delta},\bm{\Gamma}_{u \delta}$, $\bm{\Gamma}_{x x}$, $\bm{\Gamma}_{ux}$ and $\Gamma_{uu}$ as
	\begin{align}
	\mathbf{Q}_{\textnormal{d},i+1} &= \frac{1}{N} \bigg(\Delta^2 \bm{\Gamma}_{\delta \delta}+\Delta\bm{\Gamma}_{x \delta}+\Delta\bm{\Gamma}_{x \delta}^\top+\bm{\Gamma}_{xx} \notag \\
	&- \begin{bmatrix}
	\Delta\bm{\Gamma}_{x \delta}\hspace{-0.05cm}+\hspace{-0.05cm}\bm{\Gamma}_{xx} \\ \Delta\bm{\Gamma}_{u\delta}\hspace{-0.05cm}+\hspace{-0.05cm}\bm{\Gamma}_{ux} 
	\end{bmatrix}^\top \hspace{-0.08cm}
	\begin{bmatrix}
	\bm{\Gamma}_{xx} & \hspace{-0.05cm}\bm{\Gamma}_{ux}^\top \\
	\bm{\Gamma}_{ux} & \hspace{-0.05cm}\Gamma_{uu}
	\end{bmatrix}^{-1} \hspace{-0.08cm}\begin{bmatrix}
	\Delta\bm{\Gamma}_{x \delta}\hspace{-0.05cm}+\hspace{-0.05cm}\bm{\Gamma}_{xx} \\ \Delta\bm{\Gamma}_{u\delta}\hspace{-0.05cm}+\hspace{-0.05cm}\bm{\Gamma}_{ux} 
	\end{bmatrix} \bigg). \notag
	\end{align}
	However, replacing the following matrix identities above
	\small
	\begin{equation}
	\bm{\Gamma}_{\hspace{-0.04cm}xx} \hspace{-0.09cm}= \hspace{-0.1cm}\begin{bmatrix}
	\bm{\Gamma}_{\hspace{-0.04cm}xx} \\ \bm{\Gamma}_{\hspace{-0.04cm}ux} 
	\end{bmatrix}^{\hspace{-0.06cm}\top}\hspace{-0.12cm}
	\begin{bmatrix}
	\bm{\Gamma}_{\hspace{-0.04cm}xx} & \hspace{-0.05cm}\bm{\Gamma}_{\hspace{-0.04cm}ux}^\top \\
	\bm{\Gamma}_{\hspace{-0.04cm}ux} & \hspace{-0.05cm} \Gamma_{\hspace{-0.04cm}uu}
	\end{bmatrix}^{\hspace{-0.04cm}-\hspace{-0.02cm}1} \hspace{-0.11cm}\begin{bmatrix}
	\bm{\Gamma}_{\hspace{-0.04cm}xx} \\ \bm{\Gamma}_{\hspace{-0.04cm}ux} 
	\end{bmatrix}\hspace{-0.02cm}, \hspace{0.04cm} \bm{\Gamma}_{\hspace{-0.04cm}x\delta} \hspace{-0.09cm}= \hspace{-0.1cm}\begin{bmatrix}
	\bm{\Gamma}_{\hspace{-0.04cm}xx} \\ \bm{\Gamma}_{\hspace{-0.04cm}ux} 
	\end{bmatrix}^{\hspace{-0.06cm}\top}\hspace{-0.12cm}
	\begin{bmatrix}
	\bm{\Gamma}_{\hspace{-0.04cm}xx} & \hspace{-0.05cm}\bm{\Gamma}_{\hspace{-0.04cm}ux}^\top \\
	\bm{\Gamma}_{\hspace{-0.04cm}ux} & \hspace{-0.05cm} \Gamma_{\hspace{-0.04cm}uu}
	\end{bmatrix}^{\hspace{-0.04cm}-\hspace{-0.02cm}1} \hspace{-0.11cm} \begin{bmatrix}
	\bm{\Gamma}_{\hspace{-0.04cm}x\delta} \\ \bm{\Gamma}_{\hspace{-0.04cm}u\delta} 
	\end{bmatrix}\hspace{-0.06cm}, \notag
	\end{equation}
	\normalsize
	we conclude that
	\begin{equation}
	\mathbf{Q}_{\textnormal{in},i\hspace{-0.02cm}+\hspace{-0.02cm}1} \hspace{-0.09cm}=\hspace{-0.09cm} \frac{\mathbf{Q}_{\textnormal{d},i\hspace{-0.02cm}+\hspace{-0.02cm}1}}{\Delta} \hspace{-0.09cm}=\hspace{-0.09cm} \frac{\Delta}{N} \hspace{-0.1cm}\left(\hspace{-0.03cm}\bm{\Gamma}_{\delta \delta}\hspace{-0.03cm} -\hspace{-0.03cm} \begin{bmatrix}
	\bm{\Gamma}_{x \delta} \\ \bm{\Gamma}_{u\delta} 
	\end{bmatrix}^\top \hspace{-0.05cm}
	\begin{bmatrix}
	\bm{\Gamma}_{xx} & \hspace{-0.05cm}\bm{\Gamma}_{ux}^\top \\
	\bm{\Gamma}_{ux} & \hspace{-0.05cm} \Gamma_{uu}
	\end{bmatrix}^{\hspace{-0.03cm}-\hspace{-0.02cm}1} \hspace{-0.05cm} \begin{bmatrix}
	\bm{\Gamma}_{x \delta} \\ \bm{\Gamma}_{u \delta}
	\end{bmatrix} \hspace{-0.03cm}\right)\hspace{-0.03cm}, \notag
	\end{equation}
	which is what we wanted to prove. \hspace{3cm}\hfill \qed

\balance
\bibliography{references}
\end{document}